\documentclass[10pt]{IEEEtran}
\usepackage{url}
\usepackage[utf8]{inputenc}
\usepackage{xcolor}
\usepackage{amsmath}
\usepackage{amssymb}
\usepackage{xfrac}

\usepackage[acronyms,nonumberlist,nopostdot,nomain,nogroupskip]{glossaries}
\usepackage{tablefootnote}
\usepackage{booktabs}

\usepackage{tabularx}

\usepackage{tikz}
\usepackage{pgfplots}
\pgfplotsset{compat=newest}
\pgfplotsset{plot coordinates/math parser=false}
\newlength\fheight
\newlength\fwidth
\usetikzlibrary{plotmarks,patterns,decorations.pathreplacing,backgrounds,calc,arrows,arrows.meta,spy,matrix}
\usepgfplotslibrary{patchplots,groupplots}
\usepackage{tikzscale}
\usepackage{hyperref}

\newif\ifexttikz
\exttikzfalse

\ifexttikz
	\usetikzlibrary{external}
\fi

\usepackage{multirow}
\usepackage{tkz-kiviat}

\usepackage{subcaption}
\usepackage{caption}

\usepackage{mathtools}

\newacronym{3gpp}{3GPP}{3rd Generation Partnership Project}
\newacronym{adc}{ADC}{Analog to Digital Converter}
\newacronym{5g}{5G}{5th generation}
\newacronym{aimd}{AIMD}{Additive Increase Multiplicative Decrease}
\newacronym{am}{AM}{Acknowledged Mode}
\newacronym{amc}{AMC}{Adaptive Modulation and Coding}
\newacronym{aqm}{AQM}{Active Queue Management}
\newacronym{awgn}{AGWN}{Additive White Gaussian Noise}
\newacronym{balia}{BALIA}{Balanced Link Adaptation}
\newacronym{bdp}{BDP}{Bandwidth-Delay Product}
\newacronym{bf}{BF}{Beamforming}
\newacronym{cc}{CC}{Congestion Control}
\newacronym{cdf}{CDF}{Cumulative Distribution Function}
\newacronym{cn}{CN}{Core Network}
\newacronym{cqi}{CQI}{Channel Quality Information}
\newacronym{cp}{CP}{Control Plane}
\newacronym{csirs}{CSI-RS}{Channel State Information - Reference Signal}
\newacronym{dc}{DC}{Dual Connectivity}
\newacronym{dce}{DCE}{Direct Code Execution}
\newacronym{dci}{DCI}{Downlink Control Information}
\newacronym{dl}{DL}{Downlink}
\newacronym{dmr}{DMR}{Deadline Miss Ratio}
\newacronym{dmrs}{DMRS}{DeModulation Reference Signal}
\newacronym{e2e}{E2E}{end-to-end}
\newacronym{ecn}{ECN}{Explicit Congestion Notification}
\newacronym{edf}{EDF}{Earliest Deadline First}
\newacronym{enb}{eNB}{evolved Node Base}
\newacronym{epc}{EPC}{Evolved Packet Core}
\newacronym{es}{ES}{Edge Server}
\newacronym{fdma}{FDMA}{Frequency Division Multiple Access}
\newacronym{fdd}{FDD}{Frequency Division Duplexing}
\newacronym[firstplural=Radio Access Technologies (RATs)]{rat}{RAT}{Radio Access Technology}
\newacronym{fs}{FS}{Fast Switching}
\newacronym{ftp}{FTP}{File Transfer Protocol}
\newacronym{gnb}{gNB}{Next Generation Node Base}
\newacronym{harq}{HARQ}{Hybrid Automatic Repeat reQuest}
\newacronym{hetnet}{HetNet}{Heterogeneous Network}
\newacronym{hh}{HH}{Hard Handover}
\newacronym{hol}{HOL}{Head-of-Line}
\newacronym{ia}{IA}{Initial Access}
\newacronym{ieee}{IEEE}{Institute of Electrical and Electronics Engineers}
\newacronym{imt}{IMT}{International Mobile Telecommunication}
\newacronym{iot}{IoT}{Internet of Things}
\newacronym{ldpc}{LDPC}{Low-Density Parity Check}
\newacronym{los}{LOS}{Line-of-Sight}
\newacronym{lte}{LTE}{Long Term Evolution}
\newacronym{m2m}{M2M}{Machine to Machine}
\newacronym{mac}{MAC}{Medium Access Control}
\newacronym{mc}{MC}{Multi-Connectivity}
\newacronym{mcs}{MCS}{Modulation and Coding Scheme}
\newacronym{mec}{MEC}{Mobile Edge Cloud}
\newacronym{mi}{MI}{Mutual Information}
\newacronym{mimo}{MIMO}{Multiple Input, Multiple Output}
\newacronym{mmwave}{mmWave}{millimeter wave}
\newacronym{mptcp}{MPTCP}{Multipath TCP}
\newacronym{mr}{MR}{Maximum Rate}
\newacronym{mss}{MSS}{Maximum Segment Size}
\newacronym{mtd}{MTD}{Machine-Type Device}
\newacronym{mtu}{MTU}{Maximum Transmission Unit}
\newacronym{nfv}{NFV}{Network Function Virtualization}
\newacronym{nlos}{NLOS}{Non-Line-of-Sight}
\newacronym{nlosv}{NLOSv}{Vehicle Non-Line-of-Sight}
\newacronym{nr}{NR}{New Radio}
\newacronym{ofdm}{OFDM}{Orthogonal Frequency Division Multiplexing}
\newacronym{pdcch}{PDCCH}{Physical Downlonk Control Channel}
\newacronym{pdcp}{PDCP}{Packet Data Convergence Protocol}
\newacronym{pdsch}{PDSCH}{Physical Downlink Shared Channel}
\newacronym{pdu}{PDU}{Packet Data Unit}
\newacronym{pf}{PF}{Proportional Fair}
\newacronym{pgw}{PGW}{Packet Gateway}
\newacronym{phy}{PHY}{Physical}
\newacronym{pbch}{PBCH}{Physical Broadcast Channel}
\newacronym[plural=\gls{mme}s,firstplural=Mobility Management Entities (MMEs)]{mme}{MME}{Mobility Management Entity}
\newacronym{prb}{PRB}{Physical Resource Block}
\newacronym{pss}{PSS}{Primary Synchronization Signal}
\newacronym{pscch}{PSCCH}{Physical Sidelink Control Channel}
\newacronym{pucch}{PUCCH}{Physical Uplink Control Channel}
\newacronym{pusch}{PUSCH}{Physical Uplink Shared Channel}
\newacronym{rach}{RACH}{Random Access Channel}
\newacronym{ran}{RAN}{Radio Access Network}
\newacronym{red}{RED}{Random Early Detection}
\newacronym{rf}{RF}{Radio Frequency}
\newacronym{rlc}{RLC}{Radio Link Control}
\newacronym{rlf}{RLF}{Radio Link Failure}
\newacronym{rrc}{RRC}{Radio Resource Control}
\newacronym{rrm}{RRM}{Radio Resource Management}
\newacronym{rr}{RR}{Round Robin}
\newacronym{rs}{RS}{Remote Server}
\newacronym{rsrp}{RSRP}{Reference Signal Received Power}
\newacronym{rss}{RSS}{Received Signal Strength}
\newacronym{rtt}{RTT}{Round Trip Time}
\newacronym{rw}{RW}{Receive Window}
\newacronym{rx}{RX}{Receiver}
\newacronym{sa}{SA}{standalone}
\newacronym{sack}{SACK}{Selective Acknowledgment}
\newacronym{sap}{SAP}{Service Access Point}
\newacronym{sc}{SC}{Single Carrier}
\newacronym{sch}{SCH}{Secondary Cell Handover}
\newacronym{scoot}{SCOOT}{Split Cycle Offset Optimization Technique}
\newacronym{sdma}{SDMA}{Spatial Division Multiple Access}
\newacronym{sinr}{SINR}{Signal to Interference plus Noise Ratio}
\newacronym{sl}{SL}{Sidelink}
\newacronym{sm}{SM}{Saturation Mode}
\newacronym{snr}{SNR}{Signal-to-Noise-Ratio}
\newacronym{son}{SON}{Self-Organizing Network}
\newacronym{ss}{SS}{Synchronization Signal}
\newacronym{srs}{SRS}{Sounding Reference Signal}
\newacronym{sss}{SSS}{Secondary Synchronization Signal}
\newacronym{tb}{TB}{Transport Block}
\newacronym{tcp}{TCP}{Transmission Control Protocol}
\newacronym{tdd}{TDD}{Time Division Duplexing}
\newacronym{tdma}{TDMA}{Time Division Multiple Access}
\newacronym{tfl}{TfL}{Transport for London}
\newacronym{tm}{TM}{Transparent Mode}
\newacronym{trp}{TRP}{Transmitter Receiver Pair}
\newacronym{tti}{TTI}{Transmission Time Interval}
\newacronym{ttt}{TTT}{Time-to-Trigger}
\newacronym{tx}{TX}{Transmitter}
\newacronym{ue}{UE}{User Equipment}
\newacronym{ul}{UL}{Uplink}
\newacronym{uml}{UML}{Unified Modeling Language}
\newacronym{um}{UM}{Unacknowledged Mode}
\newacronym{utc}{UTC}{Urban Traffic Control}
\newacronym{vm}{VM}{Virtual Machine}
\newacronym{rsrq}{RSRQ}{Reference Signal Received Quality}
\newacronym{rssi}{RSSI}{Received Signal Strength Indicator}
\newacronym{crs}{CRS}{Cell Reference Signal}
\newacronym{nsa}{NSA}{Non Stand Alone}
\newacronym{mrdc}{MR-DC}{Multi \gls{rat} \gls{dc}}
\newacronym{endc}{EN-DC}{E-UTRAN-\gls{nr} \gls{dc}}
\newacronym{5gc}{5GC}{5G Core}
\newacronym{si}{SI}{Study Item}
\newacronym{iab}{IAB}{Integrated Access and Backhaul}
\newacronym{wf}{WF}{Wired-first}
\newacronym{hqf}{HQF}{Highest-quality-first}
\newacronym{pa}{PA}{Position-aware}
\newacronym{mlr}{MLR}{Maximum-local-rate}
\newacronym{wbf}{WBF}{Wired Bias Function}
\newacronym{mib}{MIB}{Master Information Block}
\newacronym{sib}{SIB}{Secondary Information Block}
\newacronym{rnti}{RNTI}{Radio Network Temporary Identifier}
\newacronym{dft}{DFT}{Discrete Fourier Transform}
\newacronym{kpi}{KPI}{Key Performance Indicator}
\newacronym{ppp}{PPP}{Poisson Point Process}
\newacronym{v2v}{V2V}{Vehicle-to-Vehicle}
\newacronym{wave}{WAVE}{Wireless Access in Vehicular Environments}
\newacronym{udp}{UDP}{User Datagram Protocol}
\newacronym{upa}{UPA}{Uniform Planar Array}
\newacronym{fec}{FEC}{Forward Error Correction}
\newacronym{v2x}{V2X}{Vehicle-To-Everything}
\newacronym{psfch}{PSFCH}{Physical Sidelink Feedback Channel}
\newacronym{pssch}{PSSCH}{Physical Sidelink Shared Channel}
\newacronym{csma}{CSMA}{Carrier Sense Multiple Access}
\newacronym{v2n}{V2N}{Vehicle-to-Network}
\newacronym{wlan}{WLAN}{Wireless Local Area Network}
\newacronym{cav}{CAV}{Connected and Autonomous Vehicle}
\newacronym{v2i}{V2I}{Vehicle-to-Infrastructure}
\newacronym{d2d}{D2D}{Device-to-Device}
\newacronym{c-its}{C-ITS}{Connected Intelligent Transportation System}
\newacronym{fr2}{FR2}{Frequency Range 2}
\newacronym{bs}{BS}{Base Station}
\newacronym{sdu}{SDU}{Service Data Unit}
\newacronym{csi}{CSI}{Channel State Information}
\newacronym{scs}{SCS}{Subcarrier Spacing}
\newacronym{sumo}{SUMO}{Simulation of Urban MObility}
\newacronym{prr}{PRR}{Packet Reception Ratio}
\newacronym{edca}{EDCA}{Enhanced Distribution Channel Access}

\tikzstyle{startstop} = [rectangle, rounded corners, minimum width=2cm, minimum height=0.5cm,text centered, draw=black]
\tikzstyle{io} = [trapezium, trapezium left angle=70, trapezium right angle=110, minimum width=3cm, minimum height=1cm, text centered, draw=black]
\tikzstyle{process} = [rectangle, minimum width=2cm, minimum height=0.5cm, text centered, draw=black, alignb=center]
\tikzstyle{decision} = [ellipse, minimum width=2cm, minimum height=1cm, text centered, draw=black]
\tikzstyle{arrow} = [thick,<->,>=stealth]
\tikzstyle{line} = [thick,>=stealth]
\tikzstyle{darrow} = [thick,<->,>=stealth,dashed]
\tikzstyle{sarrow} = [thick,->,>=stealth]
\tikzstyle{larrow} = [line width=0.1mm,dashdotted,->,>=stealth]

\makeatletter
\def\grd@save@target#1{%
  \def\grd@target{#1}}
\def\grd@save@start#1{%
  \def\grd@start{#1}}
\tikzset{
  grid with coordinates/.style={
    to path={%
      \pgfextra{%
        \edef\grd@@target{(\tikztotarget)}%
        \tikz@scan@one@point\grd@save@target\grd@@target\relax
        \edef\grd@@start{(\tikztostart)}%
        \tikz@scan@one@point\grd@save@start\grd@@start\relax
        \draw[minor help lines] (\tikztostart) grid (\tikztotarget);
        \draw[major help lines] (\tikztostart) grid (\tikztotarget);
        \grd@start
        \pgfmathsetmacro{\grd@xa}{\the\pgf@x/1cm}
        \pgfmathsetmacro{\grd@ya}{\the\pgf@y/1cm}
        \grd@target
        \pgfmathsetmacro{\grd@xb}{\the\pgf@x/1cm}
        \pgfmathsetmacro{\grd@yb}{\the\pgf@y/1cm}
        \pgfmathsetmacro{\grd@xc}{\grd@xa + \pgfkeysvalueof{/tikz/grid with coordinates/major step x}}
        \pgfmathsetmacro{\grd@yc}{\grd@ya + \pgfkeysvalueof{/tikz/grid with coordinates/major step y}}
        \foreach \x in {\grd@xa,\grd@xc,...,\grd@xb}
        \node[anchor=north] at (\x,\grd@ya) {\pgfmathprintnumber{\x}};
        \foreach \y in {\grd@ya,\grd@yc,...,\grd@yb}
        \node[anchor=east] at (\grd@xa,\y) {\pgfmathprintnumber{\y}};
      }
    }
  },
  minor help lines/.style={
    help lines,
    gray,
    line cap =round,
    xstep=\pgfkeysvalueof{/tikz/grid with coordinates/minor step x},
    ystep=\pgfkeysvalueof{/tikz/grid with coordinates/minor step y}
  },
  major help lines/.style={
    help lines,
    line cap =round,
    line width=\pgfkeysvalueof{/tikz/grid with coordinates/major line width},
    xstep=\pgfkeysvalueof{/tikz/grid with coordinates/major step x},
    ystep=\pgfkeysvalueof{/tikz/grid with coordinates/major step y}
  },
  grid with coordinates/.cd,
  minor step x/.initial=.5,
  minor step y/.initial=.2,
  major step x/.initial=1,
  major step y/.initial=1,
  major line width/.initial=1pt,
}
\makeatother

\definecolor{desireRed}{RGB}{230,57,60}%
\definecolor{darkPurple}{RGB}{59,31,43}%
\definecolor{springGreen}{RGB}{37,223,145}%
\definecolor{queenBlue}{RGB}{69,123,157}%
\definecolor{spaceCadet}{RGB}{29,53,87}%

\usepackage[font=footnotesize]{caption}
\usepackage[font=footnotesize]{subcaption}

\makeglossaries

\begin{document}

\title{Towards Standardization of Millimeter Wave Vehicle-to-Vehicle Networks:  Open Challenges and Performance Evaluation}

\author{{{Tommaso Zugno},~\IEEEmembership{Student Member, IEEE}, {Matteo Drago},~\IEEEmembership{Student Member, IEEE}, \\{Marco Giordani},~\IEEEmembership{Member, IEEE}, {Michele Polese},~\IEEEmembership{Member, IEEE}, {Michele Zorzi},~\IEEEmembership{Fellow, IEEE}}
\thanks{Tommaso Zugno, Matteo Drago, Marco Giordani, and Michele Zorzi are with the Department of Information Engineering, University of Padova, Padova, Italy (email: \{zugnotom, dragomat, giordani, zorzi\}@dei.unipd.it). Michele Polese was with the Department of Information Engineering, University of Padova, Padova, Italy, and is now with Northeastern University, Boston, MA, USA (email: m.polese@northeastern.edu).}}


\maketitle

\begin{abstract}
IEEE 802.11bd and 3GPP NR V2X represent the new specifications for next generation vehicular networks, exploiting new communication technologies and new spectrum, such as the \gls{mmwave} band, to improve throughput and reduce latency. In this paper, we specifically focus on the challenges that \glspl{mmwave} introduce for \gls{v2v} networking, by reviewing the  latest standard developments and the issues that 802.11bd and NR V2X will have to address for \gls{v2v} operations at \glspl{mmwave}. To the best of our knowledge, our work is the first that considers a full-stack, end-to-end approach for the design of mmWave \gls{v2v} networks, discussing open issues that span from the physical to the higher layers, and reporting the results of an end-to-end performance evaluation that highlight the potential of mmWaves for \gls{v2v} communications.
\end{abstract}

\begin{IEEEkeywords}
5G, millimeter wave, V2V, 3GPP, IEEE
\end{IEEEkeywords}

\glsresetall
\glsunset{nr}
\glsunset{ieee}
\glsunset{3gpp}

\section{Introduction}
\label{sec:intro}

Recent advances in the automotive industry have paved the way towards \glspl{cav} to promote road safety and traffic efficiency~\cite{boban2018connected}.
The potential of \glspl{cav} will be fully unleashed through \gls{v2x} wireless communications, providing connectivity to and from cellular base stations (\gls{v2i}) and among vehicles (\gls{v2v}).
Today, the two key access technologies that enable V2X communications are IEEE 802.11p and 3GPP Cellular-V2X (C-V2X) that, however, fall short of fulfilling the foreseen extreme traffic demands (e.g., in terms of very high throughput,  ultra low latency and ultra high reliability) of future vehicular services.

In this regard, different standardization activities are currently being promoted by the IEEE and the 3GPP, with the 802.11bd~\cite{TGbd.general} and NR V2X~\cite{3gpp.38.885} specifications, respectively, to overcome the limitations of current technology.
Both standards aim at boosting the wireless capacity by encompassing the possibility of using, besides traditional sub-6 GHz frequencies, the lower part of the \gls{mmwave} spectrum, which features the availability of large chunks of untapped bandwidth. This would enable data rates in the order of hundreds of megabits per second in vehicular scenarios~\cite{drago2020millicar}, improving over 3GPP C-V2X and IEEE 802.11p, which can reach -- at most -- a few tens of megabits per second~\cite{giordani2017millimeter}.
Additionally, the unique  characteristics of the \gls{mmwave} signal, including the channel sparsity and the high temporal and angular resolution, may be used for very accurate positioning of vehicles, a critical requirement for most future vehicular services~\cite{wymeersch20175g}.
However, communication at \glspl{mmwave} introduces serious challenges for the whole protocol stack and requires the maintenance of directional transmissions~\cite{giordani2017millimeter}, due to severe path and penetration losses:
even though IEEE and 3GPP research activities are in their initial stages, adequate discussion on whether (and how) standardization proposals will be able to overcome such limitations is still missing.

\begin{figure*}[t]
	\centering
	\includegraphics[width=0.7\textwidth]{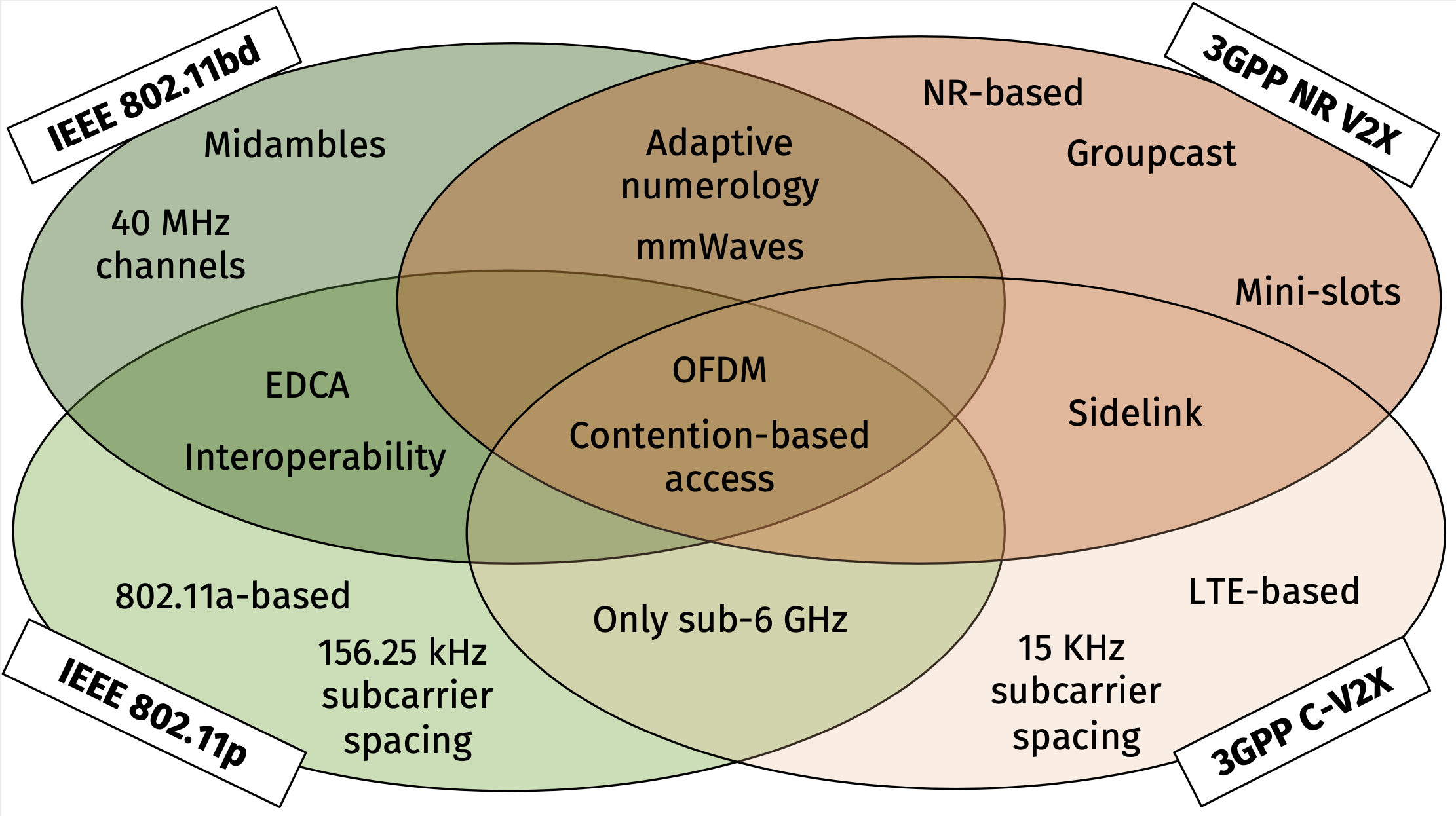}
	\caption{Common characteristics and differences of the different \gls{v2v} specifications.}
	\label{fig:standards}
\end{figure*}

Therefore, in this paper, we discuss how \gls{mmwave} operations can be efficiently integrated in IEEE 802.11bd and 3GPP NR V2X systems. Specifically, we focus on the \gls{v2v} component of these specifications, and, unlike existing literature reviewing vehicular standard developments, e.g.,~\cite{naik2019ieee}, we shed light on potential shortcomings that future releases need to overcome to fully enable V2V operations at \glspl{mmwave}.
We focus on \gls{phy}, \gls{mac}, and higher-layer design challenges, including the issues related to channel estimation, synchronization, mobility management, resource allocation and congestion and flow control.

Besides stimulating further research towards \gls{mmwave}-compliant IEEE/3GPP specifications, in this paper we validate mmWave solutions in view of the strict requirements of future vehicular systems, a research challenge that is still largely unexplored.
Therefore, we present a performance evaluation of \gls{mmwave} vehicular communications considering system-level metrics, with a novel, full-stack simulator for NR-V2X-compliant \gls{v2v} networks~\cite{drago2020millicar}, and discuss how the signal propagation at \glspl{mmwave} affects the end-to-end latency and packet reception probability for different deployment strategies.

\section{V2V Standardization Activities}
\label{sec:std}
The \gls{ieee} and the \gls{3gpp} are standardizing next-generation networks for vehicular applications with \gls{ieee} 802.11bd and 3GPP NR V2X. The two organizations aim at designing inter-operable specifications, so that these technologies can coexist in the same deployment. They are based on \gls{ofdm}, with an adaptive physical layer design, and can use both sub-6 GHz and \gls{mmwave} bands with contention-based schemes, as highlighted in Fig.~\ref{fig:standards}. Nonetheless, they also present some distinct characteristics, which are inherited from the different original designs of 3GPP and IEEE networks. NR V2X uses a sidelink for \gls{v2v} operations, which could also be scheduled, while 802.11bd is based on the 802.11 \gls{edca}. Other differences, related to the specific physical layer and signaling configurations, will be discussed in the following paragraphs.

\subsection{IEEE 802.11bd}
In March 2018, the \gls{ieee} formed the 802.11 Next Generation \gls{v2x} (NGV) Study Group, to improve the 802.11 \gls{mac} and \gls{phy} layers for \gls{v2x} communications. The current \gls{v2x} IEEE specifications, i.e., \gls{wave}, with 802.11p for the \gls{phy} and \gls{mac} layers, is derived from 802.11a - 2009, and is no longer able to guarantee the present and future needs of vehicular applications.

\begin{table*}[t!]
\caption{Millimeter wave challenges in IEEE 802.11bd and NR V2X standards.}
\label{tab:open-challenges}
\centering
\footnotesize
\renewcommand{\arraystretch}{1.2}
  \begin{tabular}{lll}
\toprule
  & Open Challenges &       Explanation     \\ \midrule
 \multirow{6}{*}{PHY Layer} & Numerology design & Longer slots lead to channel variations   \\
 & Multiple antenna arrays & Synchronization with distributed antennas   \\
  & Joint radar and communication & Based on IEEE 802.11ad (static and indoor) scenarios  \\
  & Broad/multi/groupcast communication & Directionality precludes broadcast operations   \\
  & Channel estimation & Time-varying channel hinders the use of midambles and may prevent feedback   \\
  & Synchronization & Synchronization signals need to be directional   \\ \midrule
  \multirow{3}{*}{MAC Layer} & Mobility management & Directionality complicates vehicle discovery and retransmissions \\
  & Resource allocation & CSMA strategies suffer from increased deafness \\
  & Interference management & Unscheduled and autonomous sidelink transmission prevents interference coordination   \\
  \midrule
  \multirow{3}{*}{Higher Layers} & Multi-hop  and routing & Routing is complicated by highly volatile links \\
   & Multi-RAT support & Coexistence between RATs in the same frequency band, vehicle, and/or deployment \\
   & Congestion and flow control & Suboptimal interaction between  channel variability and transport layer rate estimation \\ \midrule
 Modeling & Channel design & Effects of second order statistics, signal correlation, Doppler and fading are not characterized \\
\bottomrule
\end{tabular}
\end{table*}

The new amendment (commonly known as 802.11bd) targets communications in the 5.9 GHz band and, optionally, in the spectrum from 57 GHz to 71 GHz. Receivers implementing NGV must be able to interpret also 802.11p messages, while transmitters have to guarantee coexistence, interoperability and backward compatibility between 802.11p and 802.11bd. The goals are to reduce the \gls{e2e} latency, to increase the throughput and the communication range (up to twice those yielded by 802.11p), and to double the relative speed between vehicles (i.e., up to 500 km/h). To meet these requirements, the technology guidelines investigated so far for the sub-6 GHz band are:
\begin{itemize}
	\item the usage of \gls{ldpc} codes with midambles, i.e., specific portions of a frame in between \gls{ofdm} data symbols used to provide a better channel estimate in fast varying channels \cite{TGbd.numerologies};
	\item flexible sub-carrier spacing, with up to 40 MHz channel bandwidth.
\end{itemize}
No specifications for \glspl{mmwave} have been released yet by the IEEE, except for a proposal to upgrade part of the \gls{phy} and lower \gls{mac} layers to those designed for 11ad/11ay high data rate scenarios \cite{TGbd.ocb60}, although these standards have been designed to target indoor communications. Therefore, there is an ongoing discussion on how to address the specific challenges of this frequency range, and preliminary studies have been carried out using 802.11ad/ay.

\subsection{NR V2X}
The 3GPP has specified in Study Items for Releases 15 and 16 that C-V2X (defined specifically for LTE in Release 14, but with a forward compatible evolution path) will be extended into NR V2X, to enable next generation use cases such as vehicle platooning and advanced and remote driving, and to support high data rates for the exchange of sensor data.

The novelties investigated by the \gls{3gpp} are:
\begin{itemize}
	\item direct measurement of the \gls{sl} channel, or decoding of \gls{pscch} transmissions, to identify occupied \gls{sl} resources;
	\item multiplexing of different logical channels~\cite{3gpp.38.885}, along with the definition of the resource allocation modes 1, where the base station schedules the resources, and 2, which lets the \gls{ue} autonomously select the sidelink transmission resources. Mode 2 is the more likely candidate for an initial deployment of NR V2X, given that mode 1 would require cellular network operators to upgrade their base stations to the NR V2X specifications, with increased deployment and management costs;
	\item support of mini-slot scheduling, i.e., the possibility to immediately schedule a transmission in just a portion of the 14 \gls{ofdm} symbols specified for an NR slot, for latency-critical services;
	\item improvement of the localization accuracy of vehicles, leveraging the additional spatial and angular degrees of freedom  provided by operations at \glspl{mmwave} and the utilization of large antenna arrays;
\end{itemize}

With respect to the \gls{phy} layer numerology, no specifications have been released yet; the assumption has been to use a flexible numerology as described in 3GPP Release 15, with sub-carrier spacings of 60 and 120 KHz in Frequency Range 2 (FR2), i.e., between 24.25 and 52.6 GHz. Many other features are derived from NR.
Moreover, no further specifications have been provided about resource allocation and channel sensing at \glspl{mmwave}. Channel access schemes have not yet been specified for Release 16 and, due to lack of time until the end of the current release, NR V2X \gls{sl} enhancements will be discussed from Release 17 on.
As of December 2019, \gls{3gpp} Release 17 NR V2X  activities include (i) \gls{sl} evaluation methodology updates; (ii) low-power low-latency resource allocation enhancement, especially for mode 2; (iii) SL discontinuous reception options for broadcast, groupcast, and unicast; and (iv) support of new SL frequency bands for single-carrier operations, including FR2-specific enhancements~\cite{RP193231}.

Finally, a channel model for \gls{v2x} communication in the sub-6 GHz band (FR1) and at mmWaves (FR2) is described in \cite{3gpp.37.885}. It also features an additional \gls{nlosv} state, occurring when the \gls{los} path is blocked by vehicles, besides the \gls{nlos} state where the path is blocked by buildings.

\section{V2V Operations at MmWaves: Open Challenges}
\label{sec:cha}
As introduced in Sec.~\ref{sec:intro}, even though the standardization is moving full pace ahead towards the first V2V deployments, the use of \gls{mmwave} frequencies to support high-capacity low-latency communications introduces new challenges for the whole protocol stack which are still open for long-term research, as highlighted in the following subsections and summarized in Table~\ref{tab:open-challenges}.

\vspace*{-0.4cm}
\subsection{PHY Layer Challenges}
\label{sub:physical_layer_challenges}

\paragraph{Numerology design} 
Both 802.11bd and NR V2X \glspl{rat} support a flexible PHY frame structure, to address different QoS requirements. A longer symbol duration (i.e., a smaller subcarrier spacing) improves the communication accuracy (because the impact of noise is less relevant), but may also lead to remarkable channel variations within a slot~\cite{giordani2018tutorial}, making mmWave \gls{v2v} communications more challenging.
As a consequence, the NR V2X frame structure can be configured in a \emph{self-contained} fashion, i.e., different  sub-frames can be associated to a different numerology. In this way, it would be possible to arrange a shorter  symbol duration to support  high-data-rate low-latency applications (e.g., for cooperative perception and/or remote driving services) while a lower subcarrier spacing can be reserved for  narrowband communications to exchange basic safety information.
Moreover, 3GPP NR V2X numerology is currently based on 3GPP NR specifications for cellular scenarios, and might therefore not fit a vehicular system, due to the harsh propagation characteristics of highly mobile vehicular nodes.


\paragraph{Multiple antenna arrays}
\gls{mmwave} networks must establish directional transmissions to sustain an acceptable communication quality with beamforming.
This is  achieved using high-dimensional phased arrays, possibly placed in distributed locations~\cite{giordani2018tutorial}.
Distributed antennas improve the spectral efficiency by exploiting spatial diversity, thereby resulting in less correlated channels, but raise synchronization issues and require the design of efficient transmit power allocation and resource management mechanisms~\cite{you2010cooperative}.
In these regards,  zero-forcing and intra-block diagonalization schemes offer a good trade-off between  capacity and system complexity considering power constraints, even though more advanced studies are needed before distributed antenna solutions can be applied to vehicular networks.

\paragraph{Joint radar and communication}
The use of \glspl{mmwave} in a vehicular context is not new, with automotive radars operating in the 77 GHz spectrum. Dual-functional stacks integrating radar and \gls{v2v} communications have already been investigated in the literature~\cite{kumari2018radar}, but not combined yet in \gls{v2v} specifications.
Spectrum isolation or interference mitigation schemes typically enable their coexistence, but
a better performance would be achieved by multiplexing both sensing and data on the same waveform, thereby improving resource utilization while reducing hardware cost and size.

\paragraph{Broad/multi/groupcast communication}
Directionality may preclude
broadcast communications at mmWaves, if different directions cannot be used
simultaneously (as in analog beamforming). On the other hand, transceivers with hybrid and digital technologies can beamform towards as many directions as the number of radio-frequency chains in the phased array, thereby achieving broad/multi/groupcast communications.
Such architectures, however, are currently limited by hardware design and suffer from high energy consumption and computational complexity, which is critical considering the limited on-board resources of budget car models. To be energy efficient,  digital/hybrid beamformers will need to use appropriate  precoding techniques as well as converters with one or few bits of resolution. Discontinuous reception (DRX) modes, which enable receiving vehicles to temporarily disable their radio-frequency front end, can offer significant power savings when the traffic is intermittent, as in the case of  vehicular scenarios~\cite{shah2019power}.

\paragraph{Channel estimation}
Tracking the channel quality in multiple spatial directions will increase the channel estimation overhead at \glspl{mmwave}.
This is particularly challenging in \gls{v2v} applications, where the channel varies quickly over time, and the initial estimate may rapidly become obsolete.
Even though IEEE 802.11bd foresees the use of midambles~\cite{TGbd.numerologies} to handle channel variations, beamformed mmWave transmissions require specifically tailored channel estimation and precoding techniques.
Furthermore, the exchange of channel state information (e.g., through the new \gls{psfch} in NR V2X) needs to be timely, to avoid the feedback of stale information in scenarios with a highly variable channel (e.g., because of the increased Doppler effect at \glspl{mmwave})~\cite{giordani2017millimeter}.

\paragraph{Synchronization}
IEEE 802.11bd and 3GPP NR V2X mode 2 (a) specifications support autonomous sidelink operations with base stations~\cite{TGbd.general,3gpp.38.885}.
In this case, vehicles should maintain or acquire time and frequency synchronization with other users.
To this end, synchronization signals can be exchanged in pre-defined resource pools, even though the directional nature of the communication at \glspl{mmwave} may slow down the rate at which such information is acquired, thereby compromising robust synchronization.

\subsection{MAC Layer Challenges} 
\label{sub:mac}

The issues that mmWaves introduce at the MAC layer in \gls{v2v} scenarios stem from the lack of omnidirectional sensing and signaling, due to beamformed communications. Beamforming, indeed, introduces deafness to vehicles which are not beam-aligned, and complicates the design of channel access and neighbor discovery schemes. Moreover, these challenges add to those typical of the MAC layer in vehicular ad hoc scenarios.

\paragraph{Vehicle discovery and mobility management}
Directionality complicates an efficient and quick discovery of neighboring vehicles~\cite{giordani2017millimeter}. In the \gls{v2n} context, the base stations have fixed locations. In \gls{v2v} scenarios, instead, both endpoints move and could be within reach for just a few seconds. Therefore, 802.11bd and NR V2X signaling schemes should allow the vehicles to discover each other quickly, even when considering mmWave directional transmissions, and rapidly adapt the communication endpoint in highly mobile environments. Moreover, the volatility of the connection caused by the mmWave channel and by the mobility of vehicles makes retransmissions more complex.
MmWave systems can hence leverage automotive sensors, including Light Detection and Rangings (LiDARs) and videocameras, that gather information about the environment and classify surrounding objects: acquisitions from these sensors can then be used to reduce the overhead associated with link configuration and beam management, since, for example, the transmitter can detect the position of the receiver and estimate the optimal direction of communication.

\paragraph{Channel access and resource allocation} 
As mentioned in Sec.~\ref{sec:std}, the \gls{3gpp}
will likely introduce contention-based channel access in NR V2X (i.e., with the aforementioned mode 2), as in IEEE 802.11bd. When both specifications will be extended to mmWaves, they will need to cope with the interaction between directionality and the channel sensing schemes. The classic \gls{csma} strategies, prone to the hidden node problem even in sub-6 GHz bands, suffer from increased deafness at mmWaves. Moreover, contention avoidance messages, which broadcast the intent to occupy the channel, may not be received by every vehicle. 
Finally, the high mobility of the nodes may introduce unforeseen collisions (e.g., when a transmitting vehicle changes path) but also free up channel resources (e.g., when a vehicle moves outside the communication area). Therefore, the design of efficient uncoordinated channel access procedures in dynamic vehicular scenarios at mmWaves is even more challenging than in WLAN systems.
Notice that  the highly-volatile nature of the \gls{mmwave} channel in the vehicular scenario may create a larger response time for the \gls{amc} scheme loop at the MAC layer, hence requiring a margin to compensate for the possible outdated \gls{cqi}: this may lead to a suboptimal use of the  transmission capacity.

\paragraph{Interference management}
Directional communications at \glspl{mmwave} can isolate the users, reducing the interference and leading the network towards a noise-limited regime~\cite{giordani2018tutorial}. 
Nevertheless, the degree of isolation  depends on the density of vehicles and the level of spatial multipath, and interference may not be negligible in some deployments.
In these scenarios, interference management schemes may help improve the network capacity by scheduling transmissions to minimize the interference.
For example, the infrastructure-based and ad hoc deployments can be mixed to allow the network and the vehicles to coordinate and decide which resources should be  blanked to avoid interference with \gls{v2v} communications.
For the out-of-coverage case (supported by both 802.11bd and NR V2X), instead, vehicles autonomously determine sidelink transmission resources, thus further complicating interference management.

\subsection{Higher Layer Challenges} 
\label{sub:higher_layer_challenges}

\paragraph{Multi-RAT support} 
In next-generation \gls{v2v} networks, different technologies will coexist in the same vehicle and deployment, using the same or different frequency bands. For example, multi-connectivity techniques, that combine sub-6 GHz and \gls{mmwave} bands, could provide additional robustness to \gls{v2v} operations. The different \glspl{rat} from the 3GPP and IEEE should therefore be aware of each other, possibly with a user-plane integration at some layer. This integration can be exploited to efficiently disseminate the information over the different \glspl{rat}, to combine the benefits of complementary technologies, and make up for the limitations of a \gls{mmwave} standalone system.

\paragraph{Multi-hop communications and routing} 
Multi-hop relaying schemes can extend the limited mmWave range for \gls{v2v}. In particular, far-away vehicles may be interested in communicating through other vehicles that act as relays. \gls{v2v} network operations will have to cope with efficient routing and successful delivery of packets in networks with highly volatile links, exacerbating the issues that traditionally affect vehicular ad hoc networks~\cite{boban2018connected}. While routing is generally performed at the network layer, for such challenging scenarios a cooperation with the 3GPP and IEEE stacks could enable faster routing updates, based on continuous and prompt refresh of the links available as next hops.

\paragraph{Congestion and flow control} 
Communication in \gls{v2v} scenarios will be mostly bursty and among two peer vehicles, exploiting a massive amount of bandwidth in the \gls{mmwave} bands. For such short flows, TCP may not be needed, and could actually worsen the performance. The congestion window growth could indeed throttle the rate available at the application layer. With multi-hop communications and longer flows, instead, congestion control is needed. In this case, however, the available congestion control algorithms may suffer from the suboptimal interaction between the channel variability and the rate estimation at the transport layer, with consequent high latency and low resource utilization. Cross-layer solutions, in which the transport protocol is aware of the actual performance of the wireless \gls{rat}, would allow higher layers to quickly adapt and use the optimal operating mode for single- and multi-hop scenarios.
Finally, another challenge is how to provide reliability at the transport layer, e.g., through retransmissions, network coding or other \gls{fec} schemes.


\begin{figure*}[!t]
  \begin{subfigure}[t]{0.48\textwidth}
  	\centering
   \includegraphics[width=\textwidth]{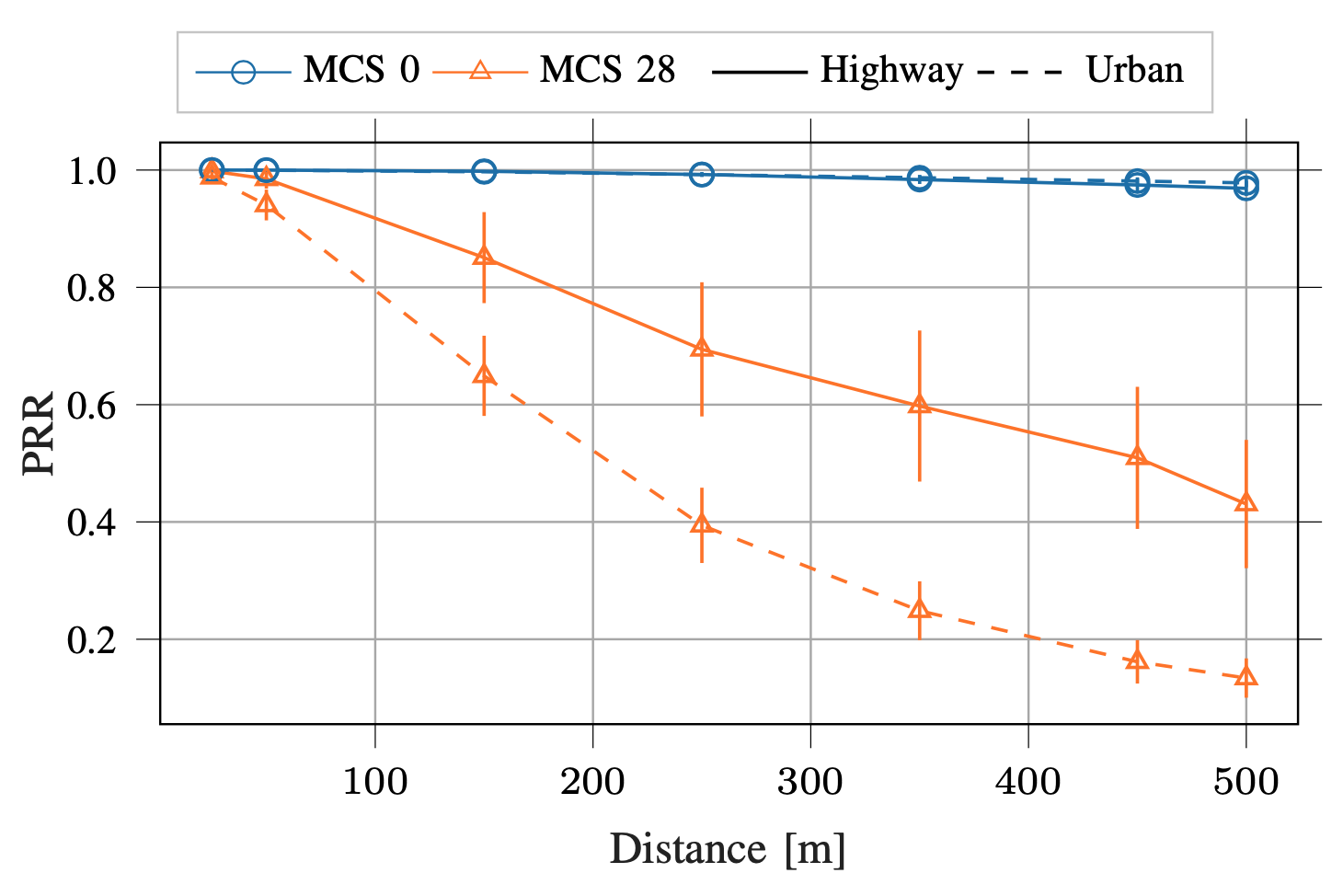}
  	\caption{Packet Reception Ratio.}
    \label{fig:hUCprr}
  \end{subfigure}
  \hfill%
\begin{subfigure}[t]{0.48\textwidth}
	\centering
 \includegraphics[width=\textwidth]{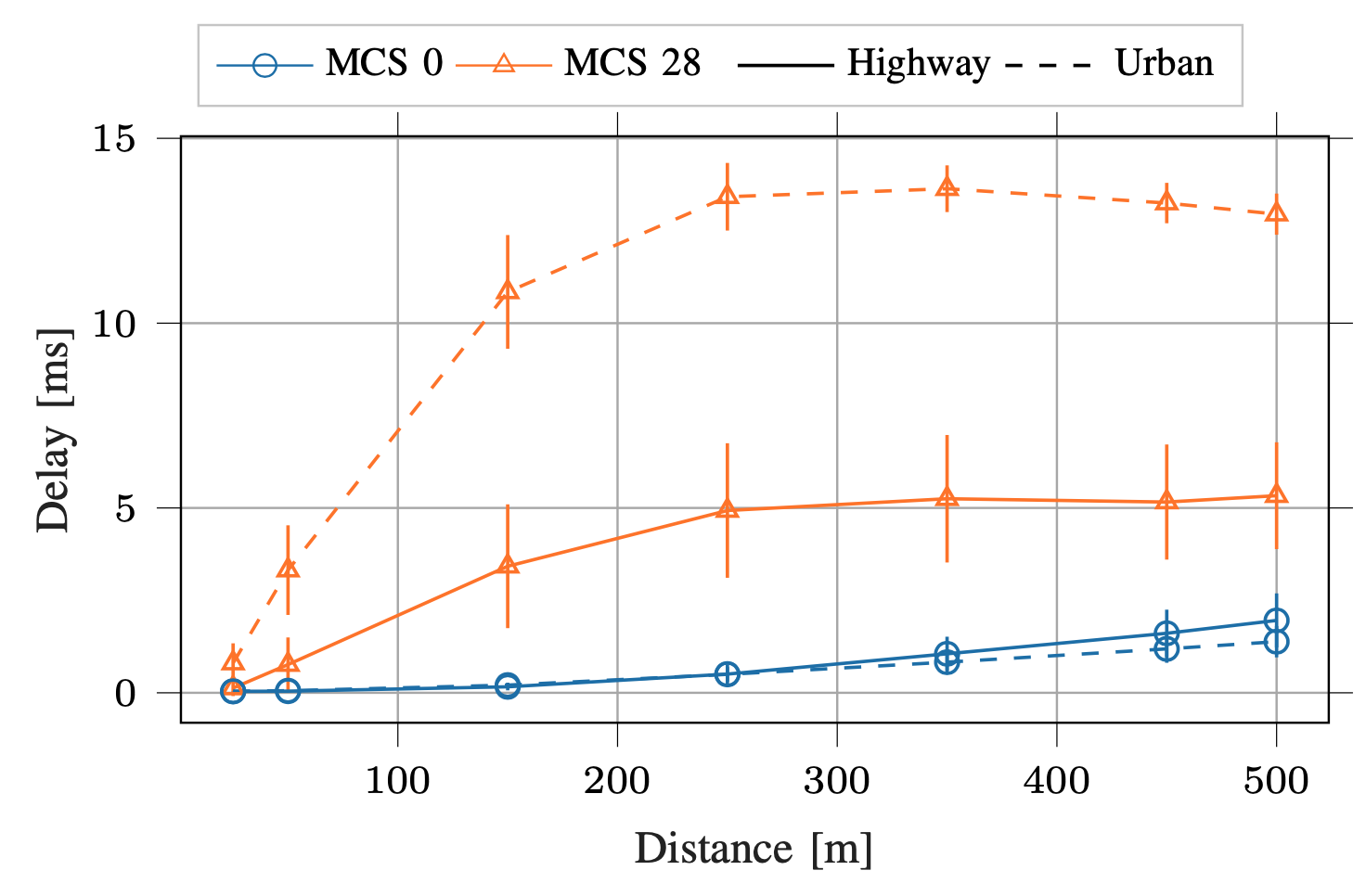}
	\caption{Delay.}
  \label{fig:hUCdelay}
\end{subfigure}%
\caption{PRR and delay vs. the relative distance between the two vehicles, for different values of the MCS.}
\label{fig:highwayUrbanComp}
\end{figure*}

\vspace{-0.3cm}
\subsection{Modeling Challenges} 

Accurate channel and protocol stack modeling at \glspl{mmwave} is an essential step towards  proper vehicular protocol design and performance characterization.
The 3GPP has specified how to characterize \gls{mmwave} propagation for NR V2X in~\cite{3gpp.37.885}, without, however, investigating second order statistics (e.g.,  spatio-temporal correlation). This prevents the applicability of existing models to dynamic environments.
Additionally, the effect of the correlation among signals in a multipath environment, e.g., the role played by reflection from adjacent vehicles, is currently underestimated.
The impact of Doppler and fading, which is critical at high frequencies, has also not yet been numerically characterized. In this sense, new measurements and the usage of ray tracing techniques could provide further insights on the performance of \gls{v2v} communications at \glspl{mmwave}, together with full-stack simulations~\cite{drago2020millicar} and real-world experiments. 





\vspace*{-0.2cm}
\section{End-to-end Performance Evaluation}
\label{sec:perf}
In this section, we present a preliminary end-to-end performance evaluation of \gls{v2v} scenarios at \glspl{mmwave}, which studies how the unique propagation characteristics experienced at high frequencies impacts vehicular environments.

\vspace*{-0.3cm}
\subsection{System Model}
The results are based on a simulation campaign using MilliCar~\cite{drago2020millicar}, an open source ns-3 module for \gls{mmwave} \gls{v2v} networks developed by the University of Padova\footnote{https://github.com/signetlabdei/millicar/}.
MilliCar features \gls{phy} and \gls{mac} layers based on the  \gls{3gpp} \gls{nr} \gls{v2x} specifications, and the \gls{3gpp} antenna and channel models for above-6-GHz \gls{v2v}~links~\cite{3gpp.37.885}.
The integration with ns-3 enables full-stack simulations of devices communicating through the \gls{nr} sidelink interface, thus assessing the system performance from an end-to-end perspective, including the effect introduced by the \gls{rlc} and \gls{pdcp} layers, as well as by the TCP/IP protocol stack.

In our simulations, we consider a platoon composed of two vehicles moving along the same lane at a constant speed of 20~m/s, one behind the other.
They communicate through the \gls{nr} sidelink interface, with a bandwidth of 100 MHz, and carriers at 28 GHz and 60 GHz. The latter is part of a large chunk of unlicensed spectrum, which could ease regulations for an initial \gls{mmwave} \gls{v2v} deployment.
Both vehicles are equipped with a \gls{upa}  with 16 antenna elements and use numerology index 2, i.e., the subcarrier spacing is $60$~kHz and the slot period is $0.25$~ms.
The front vehicle transmits \gls{udp} packets of size $100$~Bytes at a constant inter-packet interval of $1$~ms.
We evaluate the system performance in different propagation scenarios, i.e., urban and highway, using different \gls{mcs} configurations, through a Monte Carlo approach where metrics of interest are averaged over multiple independent simulations.
For a more complete description of the MilliCar module and the simulation parameters, we refer the interested reader to our previous work~\cite{drago2020millicar}.

\subsection{Performance Results}
Fig.~\ref{fig:highwayUrbanComp} shows the average \gls{prr} (Fig.~\ref{fig:hUCprr}) and delay (Fig.~\ref{fig:hUCdelay}) achieved at the application layer by varying the relative distance between vehicles.
We consider two different \gls{mcs} configurations, i.e., $4$ QAM with code rate $0.08$ (MCS 0, in blue) and $64$ QAM with code rate $0.92$ (MCS 28, in orange), and compare the performance achieved in urban (dashed lines) and highway (solid lines) environments.

Fig.~\ref{fig:hUCprr} highlights that the more robust \gls{fec} scheme and the lower modulation order provided by \gls{mcs}~0 ensure very reliable operations (PRR $\simeq 1$) even at $500$~m, both in urban and highway scenarios.
Conversely, \gls{mcs}~28 provides a weaker protection against errors, hence the \gls{prr} decreases faster as the vehicles move away from each other.
With this configuration, we can notice a significant performance gap between the urban and highway environments, which reflects the characteristics of the signal propagation in the two cases.
Indeed, in the highway scenario the signal usually propagates in free space, thus ensuring a high \gls{los} probability. In the urban case, instead, the presence of obstacles results in a severe attenuation of the received power.

This trends is confirmed by Fig.~\ref{fig:hUCdelay}, which demonstrates that a higher packet loss is associated with an increase in delay. Indeed, while \gls{mcs}~0 is able to provide an average delay lower than 2~ms even at 500~m in both urban and highway environments,  for the \gls{mcs}~28 configuration the delay increases uncontrollably with the distance, with the highway scenario performing better than its urban counterpart thanks to the higher \gls{los} probability.

\begin{figure}[t]
  \centering
	\includegraphics[width=.95\columnwidth]{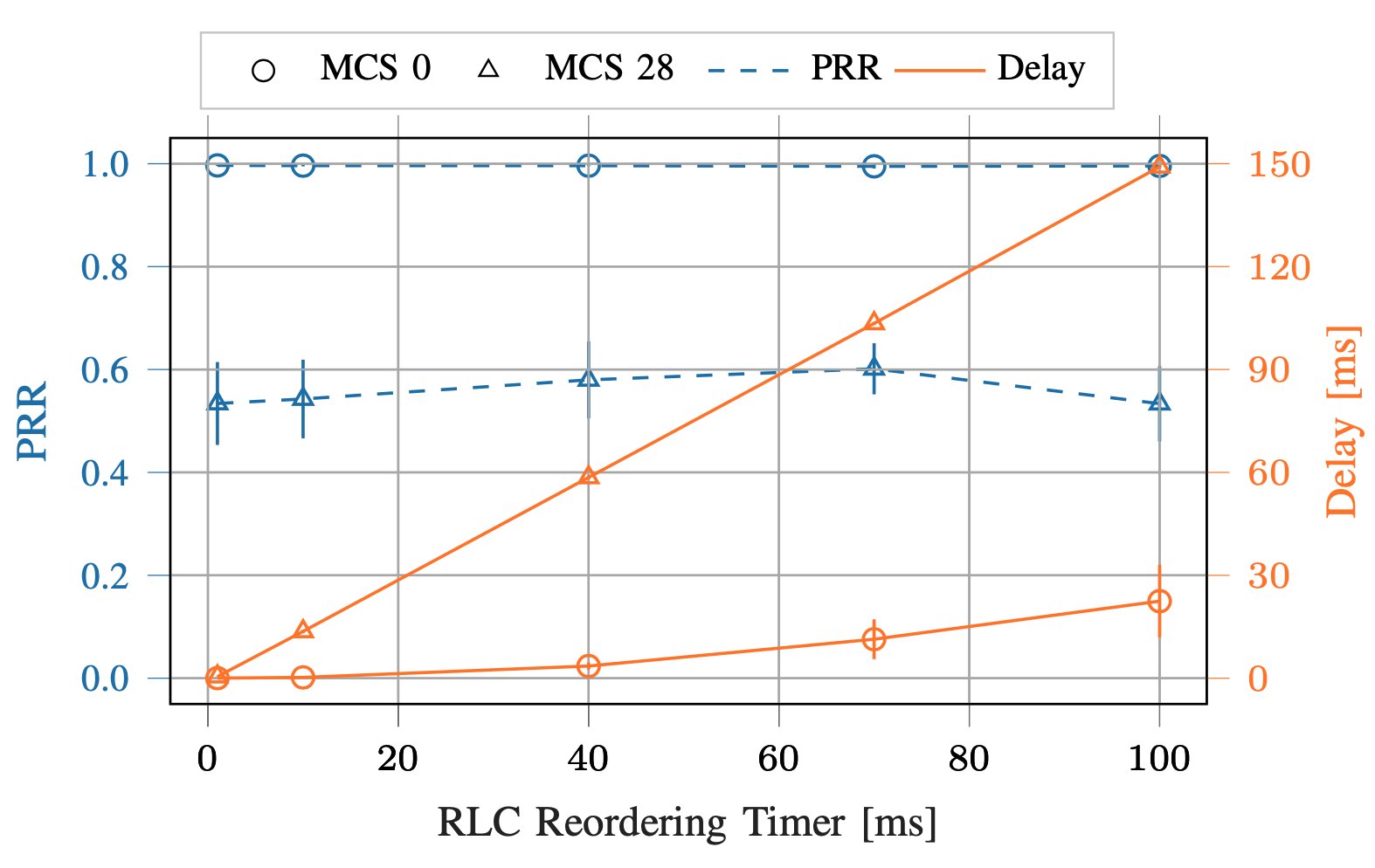}
	\caption{\gls{prr} and delay vs. the \gls{rlc} reordering timer, for different values of the MCS.}
	\label{fig:rlc_timer}
\end{figure}

We claim that the main contribution to the end-to-end delay is given by the reordering procedure at the \gls{rlc} layer. When a packet is received, it enters a reordering buffer where it has to wait for any other earlier packet (i.e., packets that were sent earlier but have not yet been received) before being forwarded to the upper layers. To avoid buffer overflow, a reordering timer is used to define the maximum waiting time after which the packet is delivered anyway. The higher the loss rate, the longer the time the received packets spend in the buffer, hence the higher the overall delay.
Along these lines, Fig.~\ref{fig:rlc_timer} investigates the impact of different values of the \gls{rlc} reordering timer on the end-to-end system performance. We jointly compare the \gls{prr} (left axis) and the average delay (right axis) when \gls{mac} layer retransmissions are not enabled. 
The results show that the delay performance deteriorates when the reordering timer is increased, showing a linear increase when using \gls{mcs}~28, while the \gls{prr} is not significantly affected (given that the absence of retransmissions makes the number of lost  packets remain constant, on average).

\begin{figure}[t]
  \centering
	\includegraphics[width=.95\columnwidth]{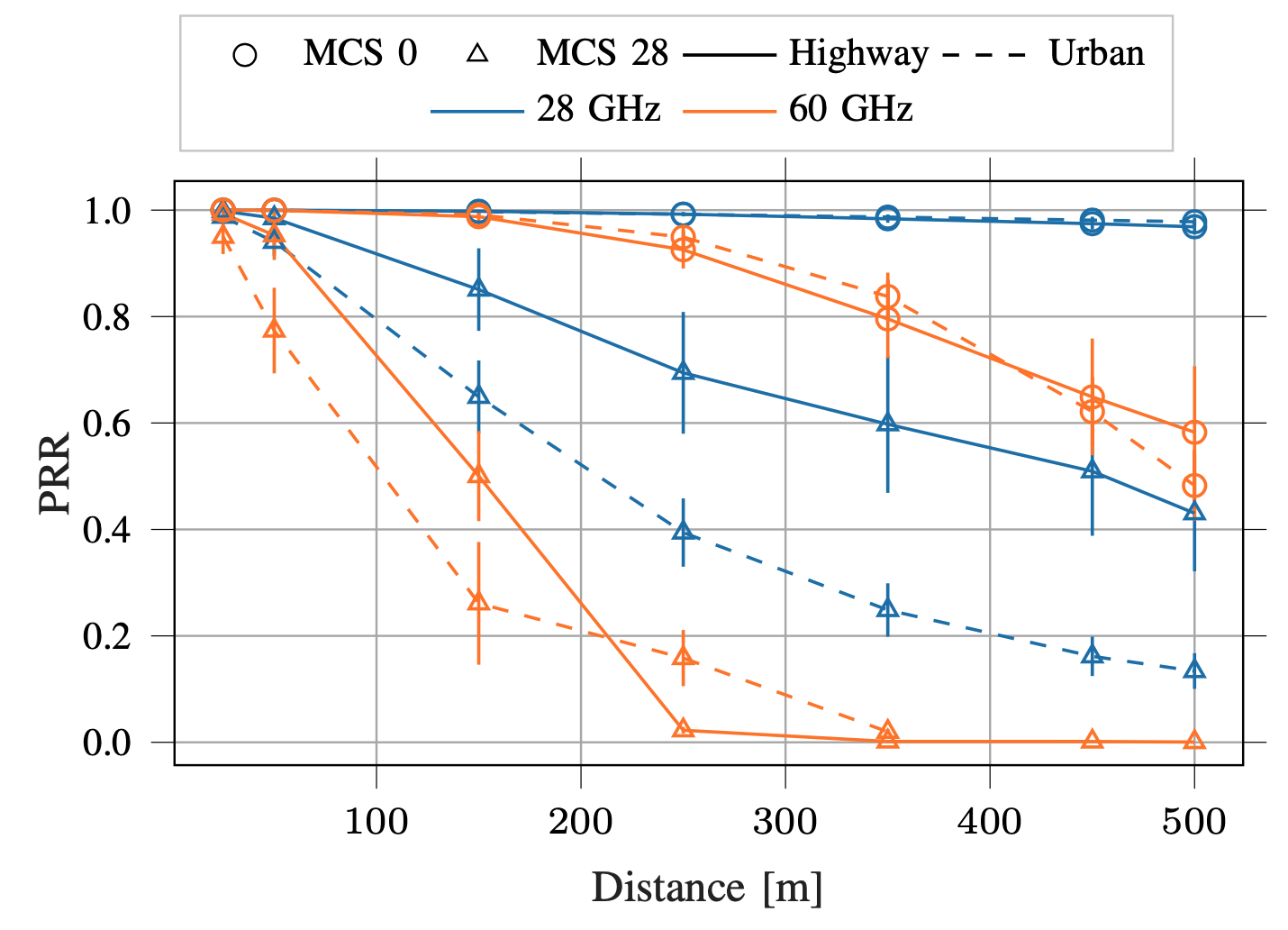}
	\caption{\gls{prr} vs. the relative distance between the two vehicles, for different propagation environments (urban and highway), and different values of the \gls{mcs} and  carrier frequency.}
	\label{fig:freq}
\end{figure}

Finally, Fig.~\ref{fig:freq} demonstrates how different values of the carrier frequency affect the \gls{prr}, as a function of the distance between the vehicles. Our results confirm that it is possible to maintain the PRR very close to 1 (a critical pre-requisite for most safety-related vehicular services)  even at large distance, when using MCS 0. 
In turn, the additional signal attenuation experienced at 60 GHz due to oxygen absorption (as severe as 15 dB/km) makes the PRR decrease significantly compared to the 28~GHz configuration. This performance gap is particularly evident when the vehicles communicate using \gls{mcs}~28 (e.g., in the highway scenario, the PRR is almost 14 times lower when considering  60 GHz transmissions at 500 m).
Finally, it is interesting to notice that, despite the oxygen absorption, the PRR  obtained in the highway environment at 60 GHz is higher than in the urban environment at 28 GHz, as long as the distance between the two vehicles is below 250 m. For larger values of  the distance, instead, the \gls{prr} drops to almost~0.


\section{Conclusions and Future Work}
This work provided an overview of the ongoing standardization activities for vehicular communications at \glspl{mmwave}, showing similarities and differences between the \gls{ieee} 802.11bd and \gls{3gpp} NR \gls{v2x} specifications.
In addition, we detailed the main challenges related to high-frequency operations considering the whole protocol stack.
Finally, we presented a preliminary end-to-end performance evaluation of a \gls{mmwave} \gls{v2v} communication system, considering different propagation environment, MCS configurations and carrier frequencies. 
As future work, we plan to evolve our end-to-end simulator following the guidelines from standardization bodies.
Furthermore, we intend to use this simulator to design novel solutions to overcome the challenges posed in Section \ref{sec:cha}, and to evaluate the overall system performance.

\bibliographystyle{IEEEtran}
\bibliography{bibl.bib}
\begin{IEEEbiographynophoto}{Tommaso Zugno}
[S’19] received his B.Sc. (2015) and M.Sc. (2018) in Telecommunication Engineering from the University of Padova, Italy. From May to October 2018 he was a Postgraduate researcher with the Department of  Information Engineering, University of Padova, under the supervision of Prof. Michele Zorzi. Since October 2018 he has been a Ph.D. student at the same university. His research focuses on protocols and architectures for 5G mmWave networks.
\end{IEEEbiographynophoto}

\begin{IEEEbiographynophoto}{Matteo Drago}
[S’19] received his B.Sc. (2016) and M.Sc. (2019) in Telecommunication Engineering from the University of Padova, Italy. Since October 2019, he has been a Ph.D. Student at the University of Padova, under the supervision of Prof. Michele Zorzi. He visited Nokia Bell Labs, Dublin, in 2018, working on QoS provisioning in 60 GHz networks. His research interests are in the study of the next generation of vehicular and millimeter-wave networks.
\end{IEEEbiographynophoto}

\begin{IEEEbiographynophoto}{Marco Giordani}
[M'20] was a Ph.D. student in Information Engineering at the University of Padova, Italy (2016-2019),  where he is now a postdoctoral researcher and adjunct professor.
He visited  NYU and TOYOTA Infotechnology Center, Inc., USA. in 2016 and 2018, respectively.
In 2018 he received the “Daniel E. Noble Fellowship Award” from the IEEE Vehicular Technology Society. His research  focuses on protocol design for 5G/6G cellular and vehicular networks.
\end{IEEEbiographynophoto}

\begin{IEEEbiographynophoto}{Michele Polese}
[M'20] is a research scientist at Northeastern University, Boston. He obtained his Ph.D. from the University of Padova, Italy, in 2020, where he also was a postdoctoral researcher and adjunct professor. He visited NYU, AT\&T Labs, and Northeastern University. His research focuses on protocols and architectures for future wireless networks.
\end{IEEEbiographynophoto}

\begin{IEEEbiographynophoto}{Michele Zorzi}
[F'07] is a Professor at the Information Engineering Department of the University of Padova, focusing on wireless communications research. He was Editor-in-Chief of IEEE Wireless Communications from 2003 to 2005, IEEE Transactions on Communications from 2008 to 2011, and IEEE Transactions on Cognitive Communications and Networking from 2014 to 2018. He served ComSoc as a Member-at-Large of the Board of Governors from 2009 to 2011, as Director of Education and Training from 2014 to 2015, and as Director of Journals from 2020 to 2021.
\end{IEEEbiographynophoto}

\end{document}